\documentclass[aps,floatfix,showpacs]{revtex4}
\usepackage{graphicx}

\begin{document}

\title{Dissipative Cylindrical Collapse of Charged Anisotropic Fluid}

\author{Sarbari Guha$^{1}$ and Ranajoy Banerji$^{2}$\footnote{RB did portions of this work during a summer program as a master's student of the Dept. of Physics and Astrophysics, University of Delhi.}}
\affiliation{$^{1}$Department of Physics, St. Xavier's College (Autonomous), 30 Mother Teresa Sarani, Kolkata, India\\
$^{2}$Saha Institute of Nuclear Physics, Kolkata, India}

\begin{abstract}
We have studied the dynamics of a cylindrical column of anisotropic, charged fluid which is experiencing dissipation in the form of heat flow, free-streaming radiation, and shearing viscosity, undergoing gravitational collapse. We calculate the Einstein-Maxwell field equations and, using the Darmois junction conditions, match the interior non-static cylindrically symmetric space-time with the exterior anisotropic, charged, cylindrically symmetric space-time. The behavior of the density, pressure and luminosity of the collapsing matter has been analyzed. From the dynamical equations, the effect of charge and dissipative quantities over the cylindrical collapse are studied. Finally, we have derived the solutions for the collapsing matter which is valid during the later stages of collapse and have discussed the significance from a physical standpoint.
\end{abstract}


\maketitle

\section{Introduction}
Gravitational collapse with realistic astronomical matter distribution is an important problem in relativistic gravity and astrophysics \cite{JM,Singh,Milne}. Over the years, there has been an extensive study of collapse of dust and fluids under gravity starting from the works of Chandrasekhar, Zwicky, Oppenheimer and Snyder \cite{Chandra,Zwicky,OS1}. Vaidya \cite{Vaidya1,Vaidya2} studied the external gravitational field of a stellar body giving out radiations. Misner and Sharp \cite{MS,Misner} studied spherically symmetric collapse. Others \cite{LSM,LH,HS,HPHPS,HDS1,Santos,Chan,BOS,BDCB,PHDMS} studied different cases of spherically symmetric fluids undergoing collapse.

Although classical considerations rule out the existence of physical objects with large amounts of charge, yet there are mechanisms which give rise to huge amount of electric charge in objects collapsing under the effect of self gravity. Rosseland \cite{Rosseland} indicated that in the stellar ensemble the atoms are strongly ionized and since the forces between the free particles should follow the inverse square law, it should be of higher order of magnitude than the residual forces acting between neutral atoms. For a star with 1.5 times solar mass and mean molecular weight 2.8, the effect of electrical forces is substantial if the star is built of heavy elements. Eddington \cite{Eddington} showed that a star has an internal electric field for which the electric potential $\phi$ depends on the gravitational potential $\psi$, the mass $m_{p}$ and charge $e$ of a proton, and a scalar parameter $\alpha$ which in turn depends on the density $n_i$ of the ions, atomic weight $A_i$ of the ions, and the effective charge $e Z_i$.

Raychaudhuri and De \cite{RD} considered the Einstein-Maxwell equations for a charged dust without imposing any special symmetry restrictions. If the magnetic field vanishes, the electric flux through any element of area bounded by particles of the dust is a constant of motion, the vorticity and electric field being orthogonal. For irrotational motion in the absence of magnetic fields, the electric field vector is orthogonal to the surfaces with constant values of $\epsilon/\rho$. They showed the impossibility of isotropic expansion and that for a charged dust in irrotational motion in absence of magnetic fields, the expansion (or contraction) cannot be shear-free. Further, the electric field and along with it the charge density would vanish if the spatial expansion were shear-free and non-vanishing. Olson and Bailyn \cite{OB} considered stars with central mass densities larger than those reasonable for a white dwarf and found that the deviations from the Chandrasekhar model were large. They found that the charge-to-mass ratio of the star was directly proportional to the average mass density. For large central mass densities, the central charge density increased and eventually produced large internal electric fields.

Bally and Harrison \cite{BH} showed that for a star of total charge $Q$ and mass $M$, the charge-to-mass ratio is given by $Q/M = G \alpha m_p / e$ and with $\alpha \sim 1$, $Q/M$ of the order of 100 coulombs per solar mass. The positive charge within a star is not automatically screened by a negatively charged atmosphere. The scale length $L$ always exceeds the Debye length $\lambda_D$ in stellar atmospheres and the interstellar medium, and both are therefore positively charged and have approximately the same ratio of charge and mass densities as stars. The Debye length $\lambda_D$ depends on the electron density $n_e$ in a gas of temperature $T$. Thus an entire galaxy can be positively charged. Even elliptical galaxies have a size that is large compared with the Debye length of their interstellar media. Oliveira and Santos \cite{OS2} have studied the junction conditions of a collapsing non-adiabatic charged body producing radiation and have observed important physical consequence due to the presence of charge. It is possible that very high electric fields may exist in strange stars with quark matter \cite{Usov,MH} under equilibrium configurations. However these do not apply to phases of intense dynamical activity with time scales of the order of (or smaller than) the hydrostatic time scale, and for which the quasistatic approximation is not reliable (e.g. the collapse of very massive stars or the quick collapse phase preceding neutron star formation).

Gravitational collapse is known to be a highly dissipative phenomenon \cite{dissipative1,dissipative2,ST}. The evolution of massive stars is characterized by dissipation due to the emission of photons or neutrinos, or both. The diffusion approximation is based on the assumption that the energy flux of radiation is proportional to the gradient of temperature. During the process of emission, the radiative transport is closer to the diffusion approximation and not to the free-streaming limit. But there are many other situations in which the mean free path of particles transporting energy are so large that the free-streaming approximation is the viable choice. Hence in a realistic model of collapse we need to consider radiative transport with both diffusion and free streaming. The study of the collapse of a strongly elongated axisymmetric body is important since such type of collapse could occur in a real astrophysical situations. Moreover, according to numerical simulations \cite{ShapTeul}, it is a possible candidate for the violation of the cosmic censorship conjecture \cite{Penrose}, and it gives insight into the hoop conjecture \cite{Thorne1}. A realistic model of collapse should also include radial heat flux.

The study of non-spherical gravitational collapse has gained in momentum following the discovery of cylindrical and plane gravitational waves. Cylindrical gravitational waves were first studied by Einstein and Rosen \cite{Rosen,ER}. Thorne \cite{Thorne2} proposed a definition of energy for systems invariant under rotations about and translations along a symmetry axis. This is the "cylindrical energy" or "C energy" which obeys the conservation law and is locally measurable. The unique static universe of Melvin \cite{Melvin} gives an absolute minimum of the C energy contained inside any cylinder. The C energy is also used to demonstrate the resistance of magnetic field lines to cylindrical gravitational collapse. Chiba \cite{Chiba} studied the case of cylindrical dust collapse. Others \cite{Nolan,GJ,PW} investigated various aspects of cylindrical collapse of counter rotating dust and rotating cylindrical shells. Hayward studied gravitational waves, black holes and cosmic strings in cylindrical symmetry \cite{hay}. Considering the most general vacuum cylindrical spacetimes, Goncalves \cite{Goncalves} presented a formal derivation of Thorne's C-energy, based on a Hamiltonian reduction approach. For the cylindrical collapse of counter-rotating dust, Goncalves and Jhingan \cite{GJ} showed that generic regular initial data could be specified for which there were no trapped surfaces in the spacetime, and a line-like singularity was inevitably developed. Di Prisco \emph{et al.} \cite{PHDMS} studied nonadiabatic charged, dissipative, spherically symmetric gravitational collapse with shear. They \cite{PHMS} also studied shear-free cylindrical gravitational collapse for an interior non-rotating fluid with anisotropic pressures and exterior vacuum Einstein-Rosen spacetime. The case of the collapse of a heat conducting charged anisotropic fluid cylinder have been studied by Sharif and Abbas \cite{SA1} and that of a charged fluid cylinder with shear viscosity by Sharif and Fatima \cite{SF}.

In this work, we have have examined the effect of charge, heat flow, radiation and shear viscosity on the gravitational collapse of a cylindrical column of fluid, which is locally anisotropic. Local anisotropy is relevant for the description of relativistic compact objects and viscous effects are important in the formation of neutron stars. The result of Raychaudhuri and De, that the shear cannot vanish in the evolution of irrotational charged dust, underlies the importance of shear in the collapse of charged fluids. After describing the gravitational source along with the corresponding physical parameters like the expansion scalar, acceleration, shear tensor and the Einstein-Maxwell field equations in the next section, we discuss the junction conditions for the exterior Vaidya metric in presence of charge in the retarded time coordinate in section III. Subsequently the dynamical equations and the solutions in presence of shear are derived in section IV. The summary of this whole exercise is presented in section V.

\section{The Interior Metric and the Field Equations}
We consider a collapsing cylinder filled with an anisotropic, charged fluid and undergoing dissipation in the form of heat flow, free-streaming radiation, and shearing viscosity, bounded by a timelike cylindrical three-surface $\Sigma$, which divides the space-time into two distinct 4-dimensional manifolds $V^+$ and $V^-$.
\subsection{The interior spacetime}
For the interior $V^-$ space-time we take the general non-static cylindrically symmetric metric in the comoving coordinates given by \cite{SF}
\begin{equation}\label{01}
ds^2_{-} = - A^2(t,r)dt^2 + B^2(t,r)dr^2 + C^2(t,r)(d\theta^2 + dz^2)
\end{equation}
In order to represent cylindrical symmetry, the range of coordinates is required to be as follows:
\begin{center}
$-\infty<t<+\infty,\quad 0\leq r <+\infty,$\\
$0\leq \theta \leq 2\pi,\quad -\infty<z<+\infty,$
\end{center}
with the coordinate labels, $x^0=t$, $x^1=r$, $x^2=\theta$ and $x^3=z$. The interior energy-momentum tensor is given, according to relativistic hydrodynamics, as
\begin{equation}\label{02}
T_{\alpha\beta}=(\mu+P_{\bot})V_{\alpha}V_{\beta}+P_{\bot}g_{\alpha\beta}
+(P_{r}-P_{\bot})\chi_{\alpha}\chi_{\beta}+V_{\alpha}q_{\beta}+V_{\beta}q_{\alpha}+ \epsilon l_{\alpha}l_{\beta}-2\eta\sigma_{\alpha\beta},
\end{equation}
where, $\mu \rightarrow$ energy density,
$P_{\bot} \rightarrow$ tangential pressure,
$P_r \rightarrow$ radial pressure,
$q^{\alpha} \rightarrow$ heat flux,
$\epsilon \rightarrow$ the radiation density,
$V^{\alpha} \rightarrow$ 4-velocity of the fluid,
$\chi^{\alpha} \rightarrow$ unit 4-velocity in the radial direction,
$l^{\alpha} \rightarrow$ a null 4-vector and $\eta \rightarrow$ coefficient of shearing viscosity $> 0$ respectively.

The shear tensor $\sigma_{\alpha\beta}$, the 4-acceleration $a_{\alpha}$  and the expansion $\Theta$ are defined as
\begin{equation}\label{03}
\sigma_{\alpha\beta}= \frac{1}{2} \left( (V_{\alpha;\beta} + V_{\beta;\alpha}) + (a_{\alpha}V_{\beta} + a_{\beta}V_{\alpha}) \right)-\frac{1}{3}\Theta(g_{\alpha\beta} + V_{\alpha}V_{\beta}),
\end{equation}
\begin{equation}\label{04}
a_{\alpha}=V_{\alpha;\beta}V^{\beta},
\end{equation}
and
\begin{equation}\label{05}
\Theta=V^{\alpha}_{ ;\alpha}.
\end{equation}

Since we have assumed comoving coordinates for the interior metric, we have
\begin{eqnarray}\label{06}
V^{\alpha}=A^{-1}\delta^{\alpha}_{0},\quad
\chi^{\alpha}=B^{-1}\delta^{\alpha}_{1}, \quad
l^{\alpha}=A^{-1}\delta^{\alpha}_0+B^{-1}\delta^{\alpha}_{1}, \quad
q^{\alpha}=B^{-1}q{\delta}^{{\alpha}}_{1},
\end{eqnarray}
such that
\begin{eqnarray}\label{07}
V^{\alpha}V_{\alpha}=-1,\quad \chi^{\alpha}\chi_{\alpha}=1,\quad
\chi^{\alpha}V_{\alpha}=0,\quad q^{\alpha}V_{\alpha}=0,\quad l^{\alpha} V_{\alpha}=-1,\quad l^{\alpha}l_{\alpha}=0.
\end{eqnarray}

In view of the equations (\ref{04}), (\ref{05}) and (\ref{06}), we obtain the acceleration and the expansion scalar as follows:
\begin{eqnarray}
  a_{\alpha} &=& \frac{A'}{A}\delta^{1}_{\alpha},\label{08} \\
  \Theta &=& \frac{1}{A}\left(\frac{\dot{B}}{B}+2\frac{\dot{C}}{C}\right).\label{09}
\end{eqnarray}

Using (\ref{03}) to (\ref{06}), we obtain the non-zero components of the shear tensor as
\begin{equation}\label{10}
\sigma_{11}=\frac{2}{\sqrt3}{B^{2}}\sigma,\quad
\sigma_{22}=\sigma_{33}=-\frac{1}{\sqrt3}{C^{2}}\sigma.
\end{equation}
The shear scalar $\sigma$ defined by
\begin{equation}\label{11}
\sigma^2 =\frac{1}{2}\sigma_{ij}\sigma^{ij}
\end{equation}
is therefore obtained as
\begin{equation}\label{12}
\sigma=\frac{1}{\sqrt{3}A}\left(\frac{\dot{B}}{B}-\frac{\dot{C}}{C}\right).
\end{equation}

As defined by Chiba \cite{Chiba}, a cylindrically symmetric spacetime may be defined locally by the existence of two commuting, spacelike, Killing vectors, such that the orthogonal space is integrable. For such a spacetime, there exist coordinates $(\theta,z)$ such that the Killing vectors are $(\xi_\theta,\xi_z)=(\partial/\partial\theta,\partial/\partial z)$. The existence of cylindrical symmetry about an axis implies that the orbits of one of these vectors are closed but those of the other is open. Each of these Killing vectors must be hypersurface orthogonal. The norms of these Killing vectors are invariants \cite{hayward}, namely the circumferential radius
\begin{eqnarray}
\nonumber
\zeta=\sqrt{\xi_\theta\cdot\xi_\theta^\flat}=\sqrt{\xi_{(2)a}\xi^{a}_{(2)}},
\end{eqnarray}
and the specific length
\begin{eqnarray}
\nonumber
\ell=\sqrt{\xi_z\cdot\xi_z^\flat}=\sqrt{\xi_{(3)a}\xi^{a}_{(3)}}
\end{eqnarray}
with $\xi_{(2)}=\partial_{\theta},~\xi_{(3)}=\partial_{z}$, under the sign convention that spatial metrics are positive definite, the dot representing contraction and the flat $\flat$ represents the covariant dual with respect to the space-time metric. The gravitational energy per specific length in a cylindrically symmetric system (also known as C-energy) as defined by Thorne \cite{Thorne2} and modified by him to render it finite in space-time, is given by
\begin{equation}\label{13}
E=\frac{1}{8}(1-l^{-2}\nabla^{a}\tilde{r}\nabla_{a}\tilde{r}),
\end{equation}
for which
\begin{eqnarray}
\tilde{r}={\zeta}l,\nonumber
\end{eqnarray}
where the areal radius is $\tilde{r}$ and $E$ is the gravitational energy per unit specific length of the cylinder.

Analogous to the Misner and Sharp energy for spherical symmetry \cite{Poisson}, the specific energy of the cylinder due to the electromagnetic field is therefore given by
\begin{equation}\label{14}
E'=\frac{l}{8}+\frac{C}{2}\left(\frac{\dot{C}^{2}}{A^{2}}
-\frac{C'^{2}}{B^{2}}\right)+\frac{s^{2}}{2C}.
\end{equation}

\subsection{Electromagnetic energy tensor and Maxwell's equations}
The electromagnetic energy-momentum tensor for the charged fluid is given by
\begin{equation}\label{15}
T^{(em)}_{\alpha\beta}=\frac{1}{4\pi}\left(F_{\alpha}^{\gamma}F_{\beta\gamma}
-\frac{1}{4}F^{\gamma\delta}F_{\gamma\delta}g_{\alpha\beta}\right).
\end{equation}
and the corresponding Maxwell's equations are
\begin{eqnarray}
F_{\alpha\beta}&=&\psi_{\beta,\alpha}-\psi_{\alpha,\beta},\label{16}\\
{F^{\alpha\beta}}_{;\beta}&=&4\pi J^{\alpha},\label{17}
\end{eqnarray}
where $F_{\alpha\beta}$ is the electromagnetic field tensor, $\psi_{\alpha}$ is the corresponding four potential and $J_{\alpha}$ is the four current density vector. Since the charge is comoving with the fluid, the charge per unit length of the cylinder is at rest with respect to the fluid and there is no magnetic field, so that the four current density is proportional to the four velocity i.e. we have
\begin{equation}\label{18}
\psi_{\alpha}=\psi{\delta^{0}_{\alpha}}=\psi(t,r)(1,0,0,0),\quad J^{\alpha}=\rho V^{\alpha},
\end{equation}
where $\psi(t,r)$ is an arbitrary function and $\rho(t,r)$ is the charge density. So the only non-zero component of the electromagnetic field tensor is
\begin{equation}\label{18a}
F_{01}=-F_{10}=-\frac{\partial \psi}{\partial r}.
\end{equation}

Thus from the Maxwell's equations we obtain
\begin{eqnarray}
\psi''-\left(\frac{A'}{A}+\frac{B'}{B}-2\frac{C'}{C}\right){\psi'}&=&{4\pi }{\rho}AB^{2},\label{19}\\
{\dot{\psi}}'-\left(\frac{\dot{A}}{A}+\frac{\dot{B}}{B}-2\frac{\dot{C}}{C}\right){\psi'}&=&0,\label{20}
\end{eqnarray}
where the first equation is for $\alpha=0$ and the second is for $\alpha=1$. Here the dot and the prime represent the partial derivatives with respect to $t$ and $r$ respectively. Integrating (\ref{19}) we obtain
\begin{equation}\label{21}
\psi'=\frac{2 sAB}{C^{2}},
\end{equation}
where
\begin{equation}\label{22}
s(r)=2{\pi}{\int^{r}_{0}}\rho BC^{2}dr
\end{equation}
is the total charge distributed per unit length of the cylinder. Equation (\ref{21}) is in conformity with the law of conservation of charge and satisfies Eq. (\ref{20}).

\subsection{The Field Equations}
We now find the field equations for this distribution of fluid. The Einstein field equations for the interior metric can be written as
\begin{equation}\label{23}
G^{-}_{\alpha\beta}=8\pi(T^{-}_{\alpha\beta}+T^{(em)^{-}}_{\alpha\beta})
\end{equation}
where $G^{-}_{\alpha\beta}$ is the Einstein tensor for the interior metric. There are five non-zero components of (\ref{23}) for the metric (\ref{01}) with energy-momentum tensor (\ref{02}), which are
\begin{eqnarray}
G^{-}_{00}=8\pi(T^{-}_{00}+T^{(em)^{-}}_{00})\nonumber
\end{eqnarray}
i.e.
\begin{equation}\label{24}
8{\pi}{(\mu+\epsilon)}A^{2}+\frac{4s^{2}{A^{2}}}{{C^{4}}}
=\frac{\dot{C}}{C}\left(2\frac{\dot{B}}{B}+\frac{\dot{C}}{C}\right)
+\left(\frac{A}{B}\right)^{2}\left(-2\frac{C''}{C}
+\frac{C'}{C}\left(2\frac{B'}{B}-\frac{C'}{C}\right)\right).
\end{equation}
Similarly,
\begin{eqnarray}
G^{-}_{01}&=&8\pi(T^{-}_{01}+T^{(em)^{-}}_{01}),\nonumber
\end{eqnarray}
which yields
\begin{equation}\label{25}
8{\pi}(q+\epsilon)AB=2\left(\frac{\dot{C'}}{C}-\frac{\dot{B}C'}{BC}-\frac{\dot{C}A'}{CA}\right).
\end{equation}
The remaining equations are
\begin{eqnarray}\label{26}
G^{-}_{11}&=&8\pi(T^{-}_{11}+T^{(em)^{-}}_{11})\nonumber\\
&=&8{\pi}\left(P_{r}+\epsilon -\frac{4}{\sqrt{3}}{\eta}{\sigma}\right)B^{2}-\frac{4s^{2}B^{2}}{C^{4}}=8{\pi}(P_{r_{eff}}+\epsilon)B^{2} -\frac{4s^{2}B^{2}}{C^{4}} \nonumber\\
&=&-\left(\frac{B}{A}\right)^{2}\left(2\frac{\ddot{C}}{C}+\left(\frac{\dot{C}}{C}\right)^{2}
-2\frac{\dot{A}\dot{C}}{AC}\right)+\left(\frac{C'}{C}\right)^{2}+2\frac{A'C'}{AC},
\end{eqnarray}
where the effective radial pressure is defined as
\begin{eqnarray}
  P_{r_{eff}} &=& P_{r}-\frac{4}{\sqrt{3}}\eta\sigma \nonumber
\end{eqnarray}
and
\begin{eqnarray}\label{27}
G^{-}_{22}&=&8\pi(T^{-}_{22}+T^{(em)^{-}}_{22})=8{\pi}\left(P_{\bot}+\frac{2}{\sqrt{3}}{\eta}{\sigma}\right)C^{2}+\frac{4s^{2}}{C^{2}}= 8{\pi}P_{{\bot}_{eff}}C^{2}+\frac{4s^{2}}{C^{2}} \nonumber\\
&=&-\left(\frac{C}{A}\right)^{2}\left(\frac{\ddot{B}}
{B}+\frac{\ddot{C}}{C}-\frac{\dot{A}}{A}\left(\frac{\dot{B}}{B}+\frac{\dot{C}}{C}\right)+
\frac{\dot{B}\dot{C}}{BC}\right)\nonumber\\
&+&\left(\frac{C}{B}\right)^{2}\left(\frac{A''}
{A}+\frac{C''}{C}-\frac{A'}{A}\left(\frac{B'}{B}-\frac{C'}{C}\right)-\frac{B'C'}
{BC}\right),
\end{eqnarray}
with the effective tangential pressure as
\begin{eqnarray}
  P_{{\bot}_{eff}} &=& P_{\bot}+\frac{2}{\sqrt3}\eta\sigma. \nonumber
\end{eqnarray}

\section{Exterior Metric and the Junction Conditions}
Exterior to the hypersurface $\Sigma$ in the 4D manifold $V^+$, we consider Vaidya's metric \cite{Vaidya2} in presence of charge in the retarded time coordinate as considered by Chao-Guang \cite{Huang}, but with a signature flip. The introduction of the retarded time coordinate removes the singularities of the original line element. Let $M(u)$ and $Q(u)$ be the mass and charge of the fluid respectively inside the hypersurface $\Sigma$, where $u$ is the retarded time coordinate. Then the exterior field in this cylindrically symmetric spacetime can be defined as
\begin{equation}\label{28}
ds^{2}_{+}=-\left(\frac{-2M(u)}{R}+\frac{Q^{2}(u)}{R^{2}_{\Sigma}}\right)du^{2}-2dRdu+R^{2}(d\theta^2+dz^2).
\end{equation}
The intrinsic metric for the hypersurface $\Sigma$ which enables a description in comoving coordinates of the interior spacetime, is given by \cite{Misner}
\begin{equation}\label{29}
(ds^{2})_{\Sigma}=-{d{\tau}}^2+R^{2}_{\Sigma}(\tau)(d\theta^{2}+dz^{2}),
\end{equation}
where ()$_{\Sigma}$ means the value of () on $\Sigma$ and $\xi^{i}\equiv(\tau,\theta,z)$ represents the coordinates on $\Sigma$, i.e.
\begin{eqnarray}
\nonumber
  (ds^{2})_{\Sigma} &=& g_{ij}d\xi^{i}d\xi^{j}.
\end{eqnarray}

To match the interior and the exterior space-time, we follow the prescription of Darmois and Israel \cite{DI} which demands:
\begin{itemize}
\item The first fundamental form must be continuous over the hypersurface $\Sigma$ i.e., the continuity of the metrics as $V^{\pm}$ approaches $\Sigma$:
\begin{equation}\label{30}
(ds^{2})_{\Sigma}=(ds^{2}_{-})_{\Sigma}=(ds^{2}_{+})_{\Sigma}.
\end{equation}
\item  The continuity of the second fundamental form. This gives the continuity of the extrinsic curvature $K_{ij}$ at the hypersurface $\Sigma$:
\begin{equation}\label{31}
[K_{ij}]=K^{+}_{ij}-K^{-}_{ij}=0.
\end{equation}
\end{itemize}

According to Eisenhart \cite{Eisenhart}, the extrinsic curvature of $\Sigma$ is given by
\begin{equation}\label{32}
K^{\pm}_{ij}=-n^{\pm}_{\sigma}\left(\frac{{\partial}^2\chi^{\sigma}_{\pm}}
{{\partial}{\xi}^i{\partial}{\xi}^j}+{\Gamma}^{\sigma}_{{\mu}{\nu}}
\frac{{{\partial}\chi^{\mu}_{\pm}}{{\partial}\chi^{\nu}_{\pm}}}
{{\partial}{\xi}^i{\partial}{\xi}^j}\right),\quad({\sigma},
{\mu},{\nu}=0,1,2,3).
\end{equation}
where $n^{\pm}_{\sigma}$ are the outward unit normal vectors to the hypersurface $\Sigma$, $\chi^{{\pm}\mu}$ are the coordinates of $V^{\pm}$.

On the $r=\textrm{constant}$ hypersurface, we have $dr = 0$. Using this condition in (\ref{01}) and comparing with (\ref{29}) keeping in mind the junction condition (\ref{30}), we get,
\begin{eqnarray}
  \frac{dt}{d\tau} &=& A(t,r_{\Sigma})^{-1},\nonumber\\
  \label{33}\\
  R_{\Sigma}(\tau) &=& C(t,r_{\Sigma}).\nonumber
\end{eqnarray}
We may also write the exterior metric (\ref{28}) as,
\begin{eqnarray}
(ds^{2}_{+})_{\Sigma}&=& -\left[\left(\frac{-2M(u)}{R_{\Sigma}}+\frac{Q^2(u)}{R_{\Sigma}^2}\right)+\frac{2dR_{\Sigma}}{du}\right]du^{2}\nonumber\\
&+&R^{2}_{\Sigma}(d\theta^2+dz^2).\label{34}
\end{eqnarray}
Now, using the junction condition (\ref{30}) and matching with the metric on the hypersurface $\Sigma$, we get
\begin{equation}\label{35}
\frac{du}{d\tau}=\left[\frac{-2M(u)}{R_{\Sigma}}+\frac{Q^2(u)}{R_{\Sigma}^2}+\frac{2dR_{\Sigma}}{du}\right]^{-1/2}.
\end{equation}
To apply the junction conditions, we require that $\Sigma$ has the same parametrisation whether it is considered as embedded in $V^+$ or in $V^-$. In the coordinates of the interior spacetime $V^-$, the bounding surface $\Sigma$ will have the equation
\begin{equation}\label{36}
f(t,r) = r - r_{\Sigma} = 0,
\end{equation}
where $r_{\Sigma}$ is a constant.

Since the vector $\partial f/\partial \chi^{\alpha}_{-}$ is orthogonal to $\Sigma$, so the unit normal vector to $\Sigma$
in the $\chi^{\alpha}_{-}$ coordinate system is,
\begin{equation}\label{37}
n^{-}_{\alpha} = [ 0,B(t,r_{\Sigma}),0,0 ].
\end{equation}

In the coordinate system of $V^+$, the equation for the surface $\Sigma$ may be written as,
\begin{equation}\label{38}
f(u,R) = R - R_{\Sigma}(u) = 0.
\end{equation}

The vector $\partial f/\partial \chi^{\alpha}_{+}$, orthogonal to the hypersurface $\Sigma$ is therefore given by,
\begin{equation}\label{39}
\frac{\partial f}{\partial \chi^{\alpha}_{+}}=\left(-\frac{d{R_{\Sigma}}}{du},1,0,0 \right).
\end{equation}

So the unit normal to $\Sigma$ in the $V^+$ coordinate system is,
\begin{equation}\label{40}
n^{+}_{\alpha}=\left(\frac{-2M(u)}{R_{\Sigma}}+\frac{Q^2(u)}{R_{\Sigma}^2}
+\frac{2dR_{\Sigma}}{du}\right)^{-1/2}\left(-\frac{d{R_{\Sigma}}}{du},1,0,0 \right).
\end{equation}

The extrinsic curvature for the hpersurface $\Sigma$ in the $V^+$ coordinates as calculated by using (\ref{32}) is given by
\begin{eqnarray}
K^{-}_{00}&=&-\left(\frac{A'}{AB}\right)_{\Sigma},\label{41}\\
K^{-}_{22}&=& K^{-}_{33}=\left(\frac{CC'}{B}\right)_{\Sigma},\label{42}\\
K_{00}^{+}&=&\left[\frac{d^{2}u}{d\tau^{2}}\left(\frac{du}{d\tau}\right)^{-1}-
\left(\frac{M}{R^2}-\frac{Q^2}{R^3}\right)\left(\frac{du}{d\tau}\right)\right]_{\Sigma}.\label{43}\\
K_{22}^{+}&=& K_{33}^{+}=\left[R\frac{dR}{d\tau}+\left(\frac{Q^2}{R}-2M\right)
\frac{du}{d\tau}\right]_{\Sigma}.\label{44}
\end{eqnarray}

On account of the continuity of the second fundamental form given by (\ref{31}), we obtain the following relations on matching (\ref{41}) to (\ref{43}) and (\ref{42}) to (\ref{44})
\begin{eqnarray}
\left[\frac{d^{2}u}{d\tau^{2}}\left(\frac{du}{d\tau}\right)^{-1}-
\left(\frac{M}{R^2}-\frac{Q^2}{R^3}\right)\left(\frac{du}{d\tau}\right)\right]_{\Sigma}
=-\left(\frac{A'}{AB}\right)_{\Sigma},\label{45}\\
\left[R\frac{dR}{d\tau}+\left(\frac{Q^2}{R}-2M\right)
\frac{du}{d\tau}\right]_{\Sigma}=\left(\frac{CC'}{B}\right)_{\Sigma}.\label{46}
\end{eqnarray}

\section{Results}
We now use the relations obtained above and simplify them to find useful results. From (\ref{35}) we have by rearranging,
\begin{equation}\label{47}
\left(\frac{du}{d\tau}\right)\left(\frac{Q^2}{R_{\Sigma}}-2M\right)=R_{\Sigma}\left(\frac{du}{d\tau}\right)^{-1} -2R_{\Sigma}\left(\frac{dR_{\Sigma}}{d\tau} \right).
\end{equation}

Putting this value in (\ref{46}) and using (\ref{33}), we have
\begin{equation}\label{48}
\left(\frac{du}{d\tau}\right)^{-1}=\left(\frac{\dot{C}}{A}+\frac{C'}{B}\right).
\end{equation}

Again using (\ref{33}) and squaring (\ref{48}) we obtain the total energy entrapped inside the surface $\Sigma$ as follows:
\begin{equation}\label{49}
M=\frac{C}{2}\left( \left(\frac{\dot{C}}{A}\right)^2 - \left(\frac{C'}{B}\right)^2 \right)+\frac{Q^2}{2C}.
\end{equation}

Taking the interior and exterior charge to be the same on the hypersurface $\Sigma$ (i.e. $Q = s$) and using (\ref{49}) and (\ref{14}), we obtain
\begin{equation}\label{50}
E'=\frac{l}{8}+M,
\end{equation}
which indicates that the difference between the two masses is equal to $l/8$, as obtained in \cite{SA1} and \cite{SF}, which is a consequence of the least unsatisfactory definition of C-energy due to Thorne \cite{Thorne2}.

Using the expressions (\ref{08}), (\ref{09}) and (\ref{12}) we can reconstruct (\ref{25}) as follows:
\begin{equation}\label{51}
4 \pi (q+\epsilon) = \frac{1}{B} \left( \frac{1}{3}(\Theta-{\sqrt3}\sigma)'-{\sqrt3}\sigma\frac{C'}{C} \right).
\end{equation}
Differentiating (\ref{48}) with respect to $\tau$ and substituting in (\ref{45}), we obtain the following expression with the help of (\ref{48}) and (\ref{33}),
\begin{equation}\label{52}
\frac{C}{A^2}\left( \frac{\ddot{C}}{C} - \frac{\dot{C}\dot{A}}{CA} \right) + \frac{C}{AB}\left(\frac{\dot{C}'}{C} - \frac{\dot{B}C'}{BC} \right) - \frac{1}{AB} \left( \frac{\dot{C}A'}{A} + \frac{C'A'}{B} \right) = \frac{1}{C^2}\left( \frac{Q^2}{C} -M \right).
\end{equation}

Using (\ref{49}), (\ref{25}) and (\ref{26}) in (\ref{52}) and rearranging terms, we arrive at the result
\begin{equation}\label{53}
q= \left( P_{r}-\frac{4}{\sqrt{3}}{\eta}{\sigma} \right) -\frac{3s^{2}}{8{\pi}{C^{4}}},
\end{equation}
on account of the fact that $Q = s$ on the hypersurface $\Sigma$. This equation gives the relation between the heat flux, radial pressure, shear viscosity and the charge per unit length of the cylinder, over the hypersurface $\Sigma$. The result shows that for an uncharged radiating fluid without any shear viscosity, the radial pressure equals the heat flux all over the
boundary of the collapsing cylinder. Equations (\ref{51}) and (\ref{53}) are generalizations over the results obtained earlier in \cite{SA1} and \cite{SF}.

The total luminosity of the collapsing matter visible to an observer at rest at infinity is \cite{OSK}
\begin{equation}\label{54}
L_{\infty} = - \left(\frac{dM}{du}\right)_{\Sigma} = - \left(\frac{dM}{dt}\frac{dt}{d\tau}\left(\frac{du}{d\tau}\right)^{-1} \right)_{\Sigma}.
\end{equation}

Differentiating (\ref{49}) with respect to $t$ and using (\ref{33}), (\ref{48}), (\ref{25}) and (\ref{26}), we obtain
\begin{eqnarray}\label{55}
L_{\infty} = 4\pi C^2 \left( \frac{\dot{C}}{A} \left( \left( P_{r}+\epsilon-\frac{4}{\sqrt{3}}{\eta}{\sigma} \right) -\frac{3s^{2}}{8{\pi}{C^{4}}} \right) + \frac{C'}{B}(q+\epsilon) \right)\left(\frac{\dot{C}}{A}+\frac{C'}{B}\right),
\end{eqnarray}
which, in view of (\ref{53}) leads us to the expression
\begin{equation}\label{56}
L_{\infty} = 4\pi \left[ C^2 (q+\epsilon) \left(\frac{\dot{C}}{A}+\frac{C'}{B}\right)^2\right]_{\Sigma}.
\end{equation}
Thus the total luminosity of the collapsing matter as visible to a distant observer, depends on the energy flux associated with the collapse. For an observer on the boundary $\Sigma$, the luminosity is \cite{GD}
\begin{equation}\label{56a}
L_{\Sigma} = - \left[ \left(\frac{du}{d\tau}\right)^2 \frac{dM}{du} \right]_{\Sigma}.
\end{equation}
The boundary redshift of the radiation emitted by the collapsing matter can be written as
\begin{equation}\label{56b}
Z_{\Sigma} = \sqrt{\frac{L_{\Sigma}}{L_{\infty}}} - 1 = \frac{du}{d\tau} - 1 = \left(\frac{\dot{C}}{A}+\frac{C'}{B}\right)^{-1} - 1
\end{equation}
Therefore the luminosity measured by an observer at rest at infinity is reduced by the redshift in comparison to
the luminosity observed on the surface of collapsing body. When
$$\left(\frac{\dot{C}}{A}+\frac{C'}{B}\right) = 0$$
the boundary redshift attains unlimited value (i.e., $Z_{\Sigma} \rightarrow \infty$ ).

\subsection{Dynamical Equations for the Collapse}
The dynamical equations for non-adiabatic charged anisotropic fluid with shear viscosity undergoing cylindrical collapse can be obtained from the Bianchi identities $(T^{\alpha\beta}+T^{{(em)}{\alpha\beta}})_{;\beta}=0$ for energy-momentum conservation. Using (\ref{02}), (\ref{06}), (\ref{07}) and (\ref{15}), we have
\begin{eqnarray}\label{57}
\left(T^{\alpha\beta}+{T^{(em)}}^{\alpha\beta}\right)_{;\beta}V_{\alpha}&=&-\frac{1}{A}(\dot{\mu}+\dot{\epsilon})
-\frac{\dot{B}}{AB}\left(\mu+P_{r}+2\epsilon-\frac{4}{\sqrt3}\eta\sigma\right) -\frac{2\dot{C}}{AC}\left(\mu+P_{\bot}+\epsilon+\frac{2}{\sqrt3}\eta\sigma\right)\nonumber\\
&-&\frac{2(q+\epsilon)}{B}\left(\frac{A'}{A}+\frac{C'}{C}\right)-\frac{1}{B}(q'+\epsilon') =0
\end{eqnarray}
and
\begin{eqnarray}\label{58}
\left(T^{\alpha\beta}+{T^{(em)}}^{\alpha\beta}\right)_{;\beta}\chi_{a}
&=&\frac{1}{B}\left(P_{r}+\epsilon-\frac{4}{\sqrt3}\eta\sigma\right)'+
\frac{A'}{AB}\left(\mu+P_{r}+2\epsilon-\frac{4}{\sqrt3}\eta\sigma\right)
+\frac{2C'}{BC}\left(P_{r}-P_{\bot}+\epsilon-{2}{\sqrt3}\eta\sigma\right)\nonumber\\
&+&\frac{1}{A}(\dot{q}+\dot{\epsilon})+\frac{2(q+\epsilon)}{A}\left(\frac{\dot{B}}{B}+\frac{\dot{C}}{C}\right) -\frac{ss'}{\pi BC^{4}}=0.
\end{eqnarray}
To discuss the dynamics of the collapsing system, it is customary to introduce the proper time derivative
\begin{equation}\label{59}
D_{T}=\frac{1}{A}\frac{\partial}{\partial{t}},
\end{equation}
and the proper radial derivative $D_{R}$ constructed  from the
circumference radius of a cylinder inside $\Sigma$
\begin{equation}\label{60}
D_{R}=\frac{1}{R'}\frac{\partial}{\partial{r}},
\end{equation}
where
\begin{equation}\label{61}
R=C.
\end{equation}
The fluid velocity for the corresponding collapse is given by
\begin{equation}\label{62}
U=D_{T}(R)=\frac{\dot{C}}{A},
\end{equation}
which must be negative to ensure collapse to occur. Defining new variable $\varepsilon=\frac{C'}{B}$ (note that $\epsilon$ and $\varepsilon$ are different quantities) and using (\ref{14}), we have
\begin{equation}\label{63}
\varepsilon=\left[U^{2}+\frac{s^{2}}{C^{2}}-\frac{2M}{C}\right]^{1/2}.
\end{equation}
Consequently, Eq. (\ref{51}) can be re-written as follows:
\begin{equation}\label{64}
4\pi (q+\epsilon) = \varepsilon \left[\frac{1}{3}D_{R}(\Theta-{\sqrt3}\sigma)-{\sqrt3}\frac{\sigma}{R}\right].
\end{equation}
The time rate of variation of the total energy inside the collapsing cylinder is given by
\begin{equation}\label{65}
D_{T}E'=-4{\pi}R^{2}\left[\left(P_{r}+\epsilon-\frac{4}{\sqrt3}\eta\sigma-\frac{1}{32{\pi}R^{2}}\right)U
+ \varepsilon(q+\epsilon)\right]+\frac{3s^{2}U}{2R^{2}}.
\end{equation}
In the case of collapse, since $(U<0)$, the coefficient of $U$ inside the square brackets, will increase the C-energy of the cylinder if $P_{r}+\epsilon-\frac{4}{\sqrt3}\eta\sigma>\frac{1}{32{\pi}R^{2}},$ i.e. the effective radial pressure is greater than a certain value. The work done by the effective radial pressure leads to the increase of C-energy. The second term in the square brackets, due to the overall negative sign, describes the outflow of energy in the form of heat flux and radiation during the collapse. Since the collapsing cylinder contains the same species of the charges, the last term will decrease the energy of the system as $\frac{3s^2}{2R^2}$ plays the role of Coulomb force of repulsion and $U<0$.

The variation of energy between the adjacent coaxial cylinders inside the fluid is given by the expression
\begin{equation}\label{66}
D_{R}E'=4{\pi}R^{2}\left({\mu}+\epsilon+\frac{U}{\varepsilon}(q+\epsilon)\right)+\frac{l}{8}
+\frac{s}{R}D_{R}s+\frac{3s^{2}}{2R^{2}}.
\end{equation}
The first term on the right hand side gives the contribution of the energy density of the element of fluid inside a cylindrical shell, along with heat flux and radiation. Since $U<0$, the factor $\frac{U}{\varepsilon}(q+\epsilon)$ decreases the energy of the system during the collapse of the cylinder. In the remaining terms, the constant $l/8$ comes from the definition of C-energy and the other term is the electromagnetic contribution. The C-energy of the cylinder at a given instant of time is then obtained by integrating (\ref{66}) from the axis to the periphery of the cylinder. The acceleration of the collapsing matter inside the hypersurface $\Sigma$ is obtained using (\ref{14}), (\ref{26}), (\ref{62}) and (\ref{63})
\begin{equation}\label{67}
D_{T}U=-\frac{1}{R^{2}}\left(E'-\frac{l}{8}\right)
-4\pi{R}\left(P_{r}+\epsilon-\frac{4}{\sqrt{3}}\eta\sigma\right)+\frac{\varepsilon A'}{AB}+\frac{5s^{2}}{2R^{3}}.
\end{equation}
Substituting for $\frac{A'}{A}$ from Eq.(\ref{67}) into Eq.(\ref{58}), we obtain the equivalent of Newton's second law of motion for the collapsing matter in the form
\begin{eqnarray}\label{68}
&&\left(\mu+P_{r}+2\epsilon-\frac{4}{\sqrt{3}}\eta\sigma\right)D_{T}U
=-\left(\mu+P_{r}+2\epsilon-\frac{4} {\sqrt{3}}\eta\sigma\right)\left[\frac{1}{R^{2}}\left(E'-\frac{l}{8}\right) +4\pi\left(P_{r}+\epsilon-\frac{4}{\sqrt{3}}\eta\sigma\right)R
-\frac{5s^{2}}{2R^{3}}\right]\nonumber\\
&&\qquad \qquad \qquad \qquad \qquad \qquad \qquad -\varepsilon\left[D_{T}(q+\epsilon)+\frac{4(q+\epsilon)U}{R}+2(q+\epsilon)\frac{1}{A}\left(\frac{\dot{B}}{B} -\frac{\dot{C}}{C}\right)\right]\nonumber\\
&&\qquad \qquad \qquad \qquad \qquad \qquad \qquad -\varepsilon^2\left[D_{R}\left(P_{r}+\epsilon-\frac{4}{\sqrt{3}}\eta\sigma\right)+2\left(P_{r}-P_{\bot}+\epsilon
-2\sqrt{3}\eta\sigma\right)\frac{1}{R}-\frac{s}{{\pi}R^{4}}D_{R}s\right],\nonumber\\
\end{eqnarray}
which can be simplified as follows:
\begin{eqnarray}
 \left(\mu+P_{r_{eff}}+2\epsilon\right)D_{T}U &=& -\left(\mu+P_{r_{eff}}+2\epsilon\right) \left[\frac{1}{R^{2}}\left(E'-\frac{l}{8}\right)+4\pi (P_{r_{eff}}+\epsilon)R
-\frac{5s^{2}}{2R^{3}}\right]\nonumber\\
&-&\varepsilon\left[D_{T}(q+\epsilon)+\frac{4(q+\epsilon)U}{R}+\frac{2(q+\epsilon)}{A}\left(\frac{\dot{B}}{B} -\frac{\dot{C}}{C}\right)\right]\nonumber\\
&-&\varepsilon^2\left[D_{R}(P_{r_{eff}}+\epsilon)+2\left(P_{r_{eff}} - P_{{\bot}_{eff}}+\epsilon\right) \frac{1}{R}-\frac{s}{{\pi}R^{4}}D_{R}s\right].\nonumber
\end{eqnarray}
Here we assume that in general $\frac{1}{A}(\frac{\dot{B}}{B}-\frac{\dot{C}}{C})\neq0$. The left hand side of (\ref{68}) represents force. The factor $(\mu+P_{r}+2\epsilon-\frac{4}{\sqrt{3}}\eta\sigma)$ represents the inertial mass density, which gives the effect of dissipation but there is no contribution of the electric charge, nor of heat flux. The remaining
term on the left hand side is acceleration. Thus, we can say that the dynamical system will evolve radially outward or inward according as $D_{T}U<0$ or $D_{T}U>0$. The terms with a negative contribution in (\ref{68}), favors the collapse while the other contribution prevents the collapse. If both of these terms cancel each other, then a condition of hydrostatic equilibrium will be encountered.

The first term on the right hand side represents the gravitational force. The factor within the first square brackets shows the effects of specific length, effective radial pressure and the electric charge on the term $(\mu+P_{r}+2\epsilon- \frac{4}{\sqrt{3}}\eta\sigma)$ representing the active gravitational mass. The second term represents the contribution due to radiation and heat flux, which will leave the system (if there is an overall negative sign) through the outward radially directed streamlines. Thus it is in the same direction of pressure and would prevent the collapse. The third term has three main contributions: the first is the effective pressure gradient which is always negative, thereby preventing the collapse. The second is the local anisotropy of the fluid which will be negative for $P_{{r}_{eff}}<P_{{\bot}_{eff}}$, in which case it will decrease the rate of collapse. The third is the electromagnetic field term. The third term contributes negatively \cite{PHDMS} if $\frac{s}{R}>D_{R}s $. Under these conditions, the term in the third square brackets, with negative sign, contributes positively by reducing the attractive nature of the force appearing on the left hand side of this equation and hence this term will prevent the gravitational collapse.

\subsection{Solution of the Field Equations}
In their work, Di Prisco \emph{et al.} \cite{PHMS} have derived the solutions for the shearfree and isotropic case of cylindrical collapse. Sharif and Abbas \cite{SA2} have found analytical solutions for charged perfect fluid cylindrical gravitational collapse. Keeping in mind the result obtained by Raychaudhuri and De \cite{RD} in the case of the evolution of irrotational charged dust, we derive the solutions in presence of shear for the evolution of charged fluids. The solution valid for the entire duration of collapse in presence of dissipation, should be of the following form:
\begin{eqnarray}\label{69}
  A(t,r) &=& A_{0}(r)f_{1}(t), \nonumber\\
  B(t,r) &=& B_{0}(r)f_{2}(t), \\
  C(t,r) &=& C_{0}(r)f_{3}(t),\nonumber
\end{eqnarray}
where $A_{0}(r)$, $B_{0}(r)$ and $C_{0}(r)$ are solutions of a static fluid having $\mu_{0}$ as the energy density and $p_{r0}$ and $p_{\bot 0}$ as the radial and tangential pressure. Rescaling the coordinate time leads to $A(t,r)=A_{0}(r)$. Then taking a cue from \cite{Chan} for the spherically symmetric case with shear, we propose solutions of the field equations (\ref{24}) to (\ref{27}) in the form
\begin{eqnarray}\label{70}
  A(t,r) &=& A_{0}(r), \nonumber\\
  B(t,r) &=& B_{0}(r), \\
  C(t,r) &=& A_{0}(r)f(t)\nonumber.
\end{eqnarray}
The expression (\ref{12}) for the shear scalar becomes
\begin{equation}\label{71}
\sigma = - \frac{1}{\sqrt{3}A_{0}}\frac{\dot{f}}{f}.
\end{equation}
The field equations (\ref{24}) to (\ref{27}) are reduced to
\begin{equation}\label{72}
8\pi(\mu+\epsilon)=8\pi\mu_{0}+\frac{\dot{f}^2}{A_{0}^2 f^2}-\frac{4s^2}{A_{0}^4 f^4},
\end{equation}
\begin{equation}\label{73}
8\pi(q+\epsilon)=0,
\end{equation}
\begin{equation}\label{74}
8\pi(P_{r}+\epsilon) = 8\pi P_{r0} -\frac{1}{A_{0}^2}\left(\frac{2\ddot{f}}{f}+\left(\frac{\dot{f}}{f}\right)^2\right)-\frac{32\pi\eta\dot{f}}{3A_{0}f} +\frac{4s^2}{A_{0}^4 f^4},
\end{equation}
\begin{equation}\label{75}
8\pi P_{\bot}=8\pi P_{\bot 0}-\frac{\ddot{f}}{A_{0}^2f}+\frac{16\pi\eta\dot{f}}{3A_{0}f}-\frac{4s^2}{A_{0}^4 f^4}
\end{equation}

where
\begin{equation}\label{76}
8\pi \mu_{0}=\frac{1}{B_{0}^2}\left[\frac{2A_{0}'B_{0}'}{A_{0}B_{0}}-\left(\frac{2A_{0}''}{A_{0}}+\frac{A_{0}'^2}{A_{0}^2}  \right)\right],
\end{equation}
\begin{equation}\label{77}
8\pi P_{r0} = \frac{3A_{0}'^2}{B_{0}^2{A_{0}}^2}
\end{equation}
and
\begin{equation}\label{78}
8\pi P_{\bot 0} = \frac{1}{B_{0}^2}\left(\frac{2A_{0}''}{A_{0}}-\frac{2A_{0}'B_{0}'}{A_{0}B_{0}} +\frac{A_{0}'^2}{A_{0}^2}\right).
\end{equation}
Equations (\ref{72}) to (\ref{75}) represent the static anisotropic fluid configuration in the limit $f(t)\rightarrow 1$. Substituting (\ref{71}), (\ref{73}) and (\ref{74}) into (\ref{53}) and assuming that $P_{r0}(r_\Sigma)=0$, we obtain the following differential equation:
\begin{equation}\label{79}
2f\ddot{f}+\dot{f}^2-\frac{a}{f^2}=0,
\end{equation}
where $a$ depends on the charge enclosed inside the cylinder and the static fluid conditions, i.e.
\begin{equation}\label{80}
a=\frac{s^2}{A_{0}^{2}}.
\end{equation}
Equation (\ref{79}) can be solved using maple program assuming that the system represents the static configuration at $t\rightarrow -\infty$, when $\dot{f}(t)\rightarrow 0$ and $f(t)\rightarrow 1$. In view of earlier works \cite{RD,BH}, we assume the charge enclosed inside the cylinder to be positive for the collapse in presence of shear. We find that the solutions represent the configuration of the collapsing matter when the luminosity of the collapsing matter as visible to a distant observer, vanishes and therefore represents the later stages of the collapse. Sample plots are shown in the following figures.
\begin{figure}[ht]
\includegraphics[scale=.3]{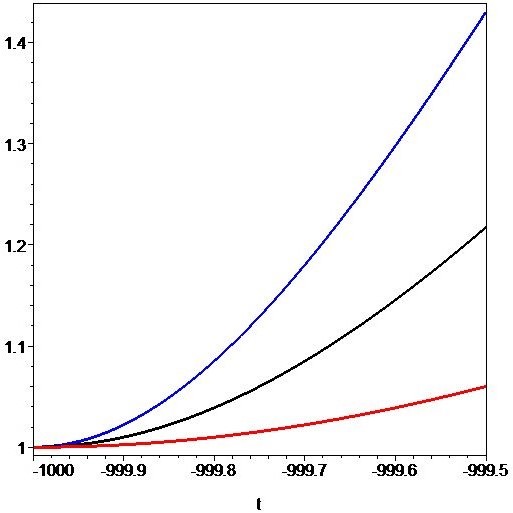}
\includegraphics[scale=.3]{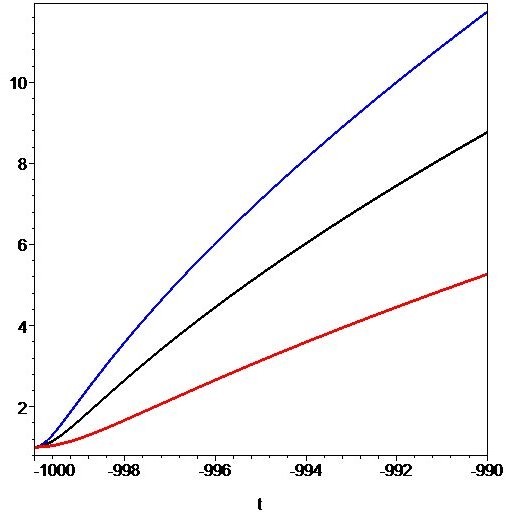}
\includegraphics[scale=.3]{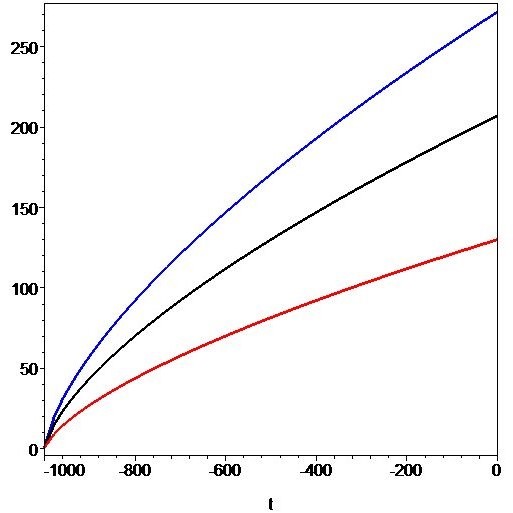}
\caption{Diagram showing the plot of $f(t)$ vs $t$ for the charged case with $a$ as squares of integers i.e $a=1$(red), $a=4$(black) and $a=9$(blue) for different ranges of time.}
\label{figureA}
\end{figure}
Analysing (\ref{79}), we find that $\dot{f}(t)$ becomes infinite as $t\rightarrow0$, which is in agreement with the trend visible in the sample plots.

\section{Summary}
Here we have formulated the general relativistic field equations for the case of dissipative cylindrical collapse in presence of heat flow, free-streaming radiation, and shear viscosity and have obtained a few results. We have derived the relation between the expansion $\Theta$, the shear $\sigma$ and the energy flowing out of the cylinder in the form of heat flux $q$ and free-streaming radiation. By employing the Darmois-Israel junction condition for the smooth matching of interior and exterior spacetimes at the boundary $\Sigma$, we have verified the relation between the specific energy of the cylinder due to the electromagnetic field and the mass of the collapsing matter. The total luminosity of the collapsing matter as visible to a distant observer, depends on the energy flux associated with the collapse. This energy flux over the hypersurface $\Sigma$ bounding the cylinder, is dependent on the effective radial pressure and the charge per unit length of the cylinder. The dynamical equations for the collapse is derived from the Bianchi identities with the help of Misner-Sharp formalism for this non-adiabatic, anisotropic and dissipative fluid and the equation for the effective Newton's second law of motion is constructed. Finally, we have derived the solution to the field equations for the given matter distribution at later stages of collapse. Under this condition the energy flux out of the boundary of the collapsing matter vanishes and the luminosity for distant observer also vanishes. The collapse is bounded by the event horizon.

As future work we are considering the solutions which represent the viscous collapsing matter from the onset of collapse to the final state of singularity and the status of the corresponding energy conditions.

\section*{Acknowledgments}
SG gratefully acknowledges IUCAA, India for an associateship. We thank Dr. Subenoy Chakraborty for some helpful discussions.


\begin{thebibliography}{}

\bibitem{JM} P. S. Joshi and D. Malafarina, Int. J. Mod. Phys. D, \textbf{20}, 2641 (2011).

\bibitem{Singh} T. P. Singh, J. Astrophys. Astr. \textbf{20}, 221 (1999).

\bibitem{Milne} E. A. Milne, Mon. Not. R. Astr. Soc. \textbf{91}, 4 (1930).

\bibitem{Chandra} S. Chandrasekhar, Mon. Not. R. Astr. Soc. \textbf{91}, 456 (1931); ibid. \textbf{95}, 207 (1935); Astrophys. J. \textbf{74}, 81 (1931); Observatory \textbf{57}, 373 (1934).

\bibitem{Zwicky} F. Zwicky, Astrophys. J. \textbf{88}, 522 (1938).

\bibitem{OS1} J.R. Openheimer and H. Snyder, Phys. Rev. \textbf{56}, 455 (1939).

\bibitem{Vaidya1} P. C. Vaidya, Current Science \textbf{12}, 183 (1943); Proc. Indian Acad. Sci. A \textbf{33}, 264 (1951); Phys. Rev. \textbf{83}, 10 (1951).

\bibitem{Vaidya2} P. C. Vaidya, Nature \textbf{171}, 260 (1953).

\bibitem{MS} C. W. Misner and D. Sharp, Phys. Rev. \textbf{136B}, 571 (1964).

\bibitem{Misner} C.W. Misner, Phys. Rev. \textbf{137B}, 1360 (1965).

\bibitem{LSM} R. W. Lindquist, R. A. Schwartz and C. W. Misner, Phys. Rev. \textbf{137B}, 1364 (1965).

\bibitem{LH} K. Lake and C. Hellaby, Phys. Rev. D \textbf{24}, 3019 (1981).

\bibitem{Santos} N.O. Santos, Mon. Not. R. Astr. Soc. \textbf{216}, 403 (1985).

\bibitem{HS} L. Herrera and N. O. Santos, Physics Reports \textbf{286}, 53 (1997).

\bibitem{HPHPS} L. Herrera, A. Di Prisco, J.L. Hernandez-Pastora and N.O. Santos, Phys. Lett. A \textbf{237}, 113 (1998).

\bibitem{HDS1} L. Herrera, G. Le Denmat and N. O. Santos, Phys. Rev. D \textbf{79}, 087505 (2009).

\bibitem{Chan} R. Chan, Mon. Not. R. Astron. Soc. \textbf{316}, 588 (2000).

\bibitem{BOS} W.B. Bonnor, A.K.G. de Oliveira and N.O. Santos, Physics Reports \textbf{181}, No. 5, 269 (1989).

\bibitem{BDCB} A. Banerjee, S. B. Dutta Choudhury, and B. K. Bhui, Phys. Rev. D \textbf{40}, 670 (1989).

\bibitem{PHDMS} A. Di Prisco, L. Herrera, G. Le Denmat, M. A. H. MacCallum and N. O. Santos, Phys. Rev. D \textbf{76}, 064017 (2007).

\bibitem{Rosseland} S. Rosseland, Mon. Not. R. Astron. Soc. \textbf{84}, 720 (1924).

\bibitem{Eddington} A. S. Eddington, \emph{Internal Constitution of the Stars}, Cambridge University Press, Cambridge, (1926).
\bibitem{RD} A. K. Raychaudhuri and U. K. De, J. Phys. A \textbf{3}, 263 (1970 ).

\bibitem{OB} E. Olson and M. Bailyn, Phys. Rev. D \textbf{13}, 2204 (1976).

\bibitem{BH} J. Bally and E.R. Harrison, Astrophys. J. \textbf{220}, 743 (1978).

\bibitem{OS2} A. K. G. De Oliveira and N. O. Santtos, Astrophys. J. \textbf{312}, 640 (1987).

\bibitem{Usov} V. Usov, Phys. Rev. D \textbf{70}, 067301 (2004).

\bibitem{MH} M. Mak and T. Harko, Int. J. Mod. Phys. D \textbf{13}, 149 (2004).

\bibitem{dissipative1} E. N. Glass, Phys. Lett \textbf{86A}, 351 (1981).

\bibitem{dissipative2} L. Herrera, A. Di Prisco, J. Martin, J. Ospino, N.O. Santos, and O. Troconis, Phys. Rev. D \textbf{69}, 084026 (2004); A. Mitra, Phys. Rev. D \textbf{74}, 024010 (2006).

\bibitem{ST} R. Sharma and R. Tikekar, Gen. Relativ. Gravit. \textbf{44}, 2503 (2012).

\bibitem{ShapTeul} S. L. Shapiro and S. A. Teukolsky, Phys. Rev. Lett. \textbf{66}, 994 (1991), S. L. Shapiro and S. A. Teukolsky, Phys. Rev. D \textbf{45}, 2006 (1992).

\bibitem{Penrose} R. Penrose, Rivista del Nuovo Cimento, Numero Speziale I, 257 (1969).

\bibitem{Thorne1} K. S. Thorne, in Magic without Magic : John Archibald Wheeler, edited by J.Klauder (Freeman, San Francisco, 1972)

\bibitem{Rosen} N. Rosen, \emph{Jubilee of Relativity Theory}, Ed. A. Mercier and M. Kervaire, Birkhauser Verlag, Basel, (1956).

\bibitem{ER} A. Einstein and N. Rosen, J. Franklin Inst. \textbf{223}, 43 (1937); N. Rosen, Bull. Research Council Israel \textbf{3}, 528 (1953).

\bibitem{Thorne2} K.S. Thorne, Phys. Rev. \textbf{138B}, 251 (1965).

\bibitem{Melvin} M. A. Melvin, Phys. Rev. \textbf{139B}, 225 (1965); M. A. Melvin, Phys. Letters \textbf{8}, 65 (1964).

\bibitem{Chiba} T. Chiba, Prog. Theor. Phys. \textbf{95}, 321 (1996).

\bibitem{Nolan} B. C. Nolan, Phys. Rev. D \textbf{65}, 104006(2002).

\bibitem{GJ} S. M. C. V. Goncalves and S. Jhingan, Int. J. Mod. Phys. D \textbf{11}, 1469 (2002).

\bibitem{PW} P. R. C. T. Pereira and A. Wang, Phys. Rev. D \textbf{62}, 124001 (2000); Erratum-ibid. \textbf{D67}, 129902 (2003); Gen. Relativ. Gravit. \textbf{32}, 2189 (2000).

\bibitem{hay} S. A. Hayward, Class. Quantum Grav. \textbf{17}, 1749 (2000).

\bibitem{Goncalves} S. M. C. V. Goncalves, Class. Quantum Grav. \textbf{20}, 37 (2003).

\bibitem{PHMS} A. Di Prisco, L. Herrera, M. A. H. MacCallum and N. O. Santos, Phys. Rev. D \textbf{80}, 064031 (2009).

\bibitem{SA1} M. Sharif and G. Abbas, Astrophys. Space Sci. \textbf{335}, 515 (2011).

\bibitem{SF} M. Sharif and S. Fatima, Gen. Relativ. Gravit. \textbf{43}, 127 (2011).

\bibitem{hayward} Here we follow the definition given by Hayward. See, \cite{hay}.

\bibitem{Poisson} E. Poisson, An Advanced Course in General Relativity, Lecture notes, Dept. of Physics, University of Guelph, 2002; \emph{A Relativist's Toolkit} (Cambridge University Press, 2004).

\bibitem{Huang} H. Chao-Guang, Acta Phys. Sin. (overseas edition) \textbf{4}, 617 (1995).

\bibitem{DI} G. Darmois, Memorial des Sciences Mathematiques (Gautheir-Villars, Paris, 1927) Fasc. 25; W. Israel, Nuovo Cimento B \textbf{44}, 1 (1966); \emph{ibid.} \textbf{48}, 463 (1966).

\bibitem{Eisenhart} L. P. Eisenhart, \emph{Riemannian Geometry} (Princeton University Press, 1949).

\bibitem{OSK} A. K. G. De Oliveira, N. O. Santtos and C. A. Kolassis, Mon. Not. R. Astron. Soc. \textbf{216}, 1001 (1985).

\bibitem{GD} S. G. Ghosh and D. W. Deshkar, Int. J. Mod. Phys. D \textbf{12}, 317 (2003); S. Nath, U. Debnath and S. Chakraborty, Astrophys. Space Sci. \textbf{313}, 431 (2008).

\bibitem{SA2} M. Sharif and G. Abbas, Jour. Phys. Soc. Japan \textbf{80}, 104002 (2011).

\end{thebibliography}
\end{document}